\documentclass{article}

\usepackage{PRIMEarxiv}

\usepackage[utf8]{inputenc} 
\usepackage[T1]{fontenc}    
\usepackage{hyperref}       
\usepackage{url}            
\usepackage{booktabs}       
\usepackage{amsfonts}       
\usepackage{nicefrac}       
\usepackage{microtype}      
\usepackage{lipsum}
\usepackage{fancyhdr}       
\usepackage{graphicx}       
\graphicspath{{media/}}     
\usepackage{multirow}   
\usepackage{makecell}   
\usepackage{caption}    
\usepackage{amsmath}    

\title{Accurate Performance Predictors for Edge Computing Applications
\thanks{This manuscript is a preprint and has been submitted for possible publication. It has not undergone peer review.}
}

\author{
  Panagiotis Giannakopoulos \\
  Eindhoven University of Technology \\
  Eindhoven, The Netherlands\\
  \texttt{p.giannakopoulos@tue.nl} \\
   \And
   Bart van Knippenberg \\
  Thermo Fisher Scientific \\
  Eindhoven, The Netherlands\\
  \texttt{bart.vanknippenberg@thermofisher.com} \\
  \And
   Kishor Chandra Joshi \\
  Eindhoven University of Technology \\
  Eindhoven, The Netherlands\\
  \texttt{k.c.joshi@tue.nl} \\
  \And
   Nicola Calabretta \\
  Eindhoven University of Technology \\
  Eindhoven, The Netherlands\\
  \texttt{n.calabretta@tue.nl} \\
  \And
   George Exarchakos \\
  Eindhoven University of Technology \\
  Eindhoven, The Netherlands\\
  \texttt{g.exarchakos@tue.nl} \\
}

\begin{document}
\maketitle

\begin{abstract}
Accurate prediction of application performance is critical for enabling effective scheduling and resource management in resource-constrained dynamic edge environments. However, achieving predictable performance in such environments remains challenging due to the co-location of multiple applications and the node heterogeneity. To address this, we propose a methodology that automatically builds and assesses various performance predictors. This approach prioritizes both accuracy and inference time to identify the most efficient model. Our predictors achieve up to 90\% accuracy while maintaining an inference time of less than 1\% of the Round Trip Time. These predictors are trained on the historical state of the most correlated monitoring metrics to application performance and evaluated across multiple servers in dynamic co-location scenarios. As usecase we consider electron microscopy (EM) workflows, which have stringent real-time demands and diverse resource requirements. Our findings emphasize the need for a systematic methodology that selects server-specific predictors by jointly optimizing accuracy and inference latency in dynamic co-location scenarios. Integrating such predictors into edge environments can improve resource utilization and result in predictable performance.
\end{abstract}


\section{Introduction}
As computing environments evolve to support increasingly diverse and time-sensitive applications, the limitations of centralized data centers have become more pronounced. These architectures often struggle to satisfy the stringent latency, reliability, and privacy requirements of real-time workloads. To overcome these challenges, edge computing has emerged as a promising alternative~\cite{shi_edge_2016}. By moving computational resources closer to data sources and end users, edge computing reduces communication distances, thereby enhancing responsiveness and lowering latency. However, this architectural shift introduces new complexities in resource management and makes performance less predictable.

Edge environments are composed of smaller geographically distributed infrastructures, in contrast to the centralized nature of traditional data centers. While co-locating multiple applications at the edge can improve resource utilization and lower operational costs, it also leads to resource contention and race conditions. This competition over limited computational and network resources results in variability in execution times, which undermines the predictability of performance~\cite{paasivaara_predictable_2020}. The situation is further complicated by the heterogeneous nature of edge nodes, which may differ in hardware configurations and software stacks~\cite{maricq_taming_nodate,duplyakin_studying_2019}. Consequently, effective planning and management of compute and network resources are essential to achieve predictable performance~\cite{fu_progress-based_2019}, maximize resource efficiency~\cite{townend_invited_2019}, and enable automated orchestration across edge infrastructures.

To address the challenges of system heterogeneity, co-located workloads, high-dimensional monitoring data, and the dynamic conditions of edge environments, predictive models capable of forecasting application execution times based on historical monitoring data have become increasingly important~\cite{wang_lass_2021}. These predictors support proactive orchestration and adaptive scheduling, allowing time-sensitive applications to meet performance requirements despite fluctuating conditions~\cite{wang_lass_2021,aslanpour_load_2024}. For deployment in edge environments, such models must strike a balance between accuracy and overhead. They must be accurate enough to deliver reliable predictions and lightweight enough to operate on resource-constrained nodes~\cite{aslanpour_load_2024}.

Developing prediction models, however, is non-trivial. System heterogeneity and co-located workloads introduce complex interdependencies that affect application behavior. Furthermore, monitoring systems typically collect hundreds of metrics at fine-grained intervals, resulting in high-dimensional data. Selecting the most relevant subset of monitoring metrics for prediction models is important to prevent overfitting and reduce inference time. Advanced filtering and feature selection techniques have been shown to improve both prediction accuracy and efficiency~\cite{zhang_eaas_2022,aslanpour_load_2024}. However, the dynamic nature of edge environments complicates the development of models that are accurate and generalizable over time~\cite{wang_lass_2021,zhang_eaas_2022}.

We investigate performance predictability using an electron microscopy (EM) workflow. We explore the Single Particle Analysis (SPA) acquisition and processing workflow which is crucial for near-real-time (i.e. sub-second timescales) instrument control. While EM is a unique domain, SPA allows us to study applications with various task types through its wide range of resource requirements (i.e. CPU, GPU, network).

We conduct experiments on a Kubernetes-managed cluster with GPUs, monitored every 200 milliseconds by Prometheus. Our methodology automatically trains and evaluates multiple models, choosing the best according to accuracy and inference time. Predictors use historical monitoring data (e.g., CPU usage, memory status, network activity) to estimate Round-Trip Time (RTT) under changing conditions. We select input features with perfCorrelate~\cite{giannakopoulos_perfcorrelate_2025}, a framework that identifies the metrics most related to RTT variability. Unlike methods that rely on a fixed set of metrics or access to application code stages, our approach trains predictors using a reduced, highly correlated set of metrics to RTT fluctuations. This ensures accuracy while minimizing overhead. Our results demonstrate how the models generalize under varying co-location scenarios and across different servers in the cluster. Furthermore, we demonstrate that factors such as the size of the historical state of the monitoring data affect the accuracy of the model, which in turn influences the decisions made by the proposed methodology. The achievements of this research are summarized as follows:
\begin{itemize}
\item We propose a comprehensive methodology that automatically generates efficient performance predictors tailored to edge environments, balancing accuracy with low-latency inference. Our predictors achieve up to 90\% accuracy in predicting application RTT prior to submission.
\item We evaluate the trade-offs between prediction accuracy and inference time across multiple machine learning models to predict application performance in near-real-time scenarios. The inference time of the predictors remains within 1\% of the RTT, ensuring minimal overhead.
\item The developed predictors are tailored for each server in the cluster and adjust dynamically to changes in co-located applications, accounting for new sources of performance variability.
\end{itemize}

The remainder of the paper is structured as follows. Section~\ref{related_work} presents the related work. Section~\ref{methodology} elaborates on the methodology for correlating monitoring metrics with performance. Section~\ref{experiment_set_up} outlines the experimental setup. Section~\ref{results} presents the experimental results. Section~\ref{conclusions} summarizes the conclusions and discusses future work.
\section{Related work} \label{related_work}
Scheduling mechanisms in cloud and edge environments are critical for efficiently allocating computational and network resources to meet application-specific requirements, such as bounded execution times. However, making optimal scheduling decisions is complex due to performance variability, which is influenced by multiple factors including the operating system (OS), hardware specifications, and application behavior~\cite{chunduri_run--run_2017}. Inconsistencies in disk, memory, network~\cite{maricq_taming_nodate}, and CPU performance~\cite{duplyakin_studying_2019} have been observed even during isolated, repeated executions. Performance fluctuations increase further when applications are co-located on the same host due to resource sharing~\cite{paasivaara_predictable_2020}. Furthermore, structural changes in applications can alter performance due to variations in dependency efficiency~\cite{laaber_evaluating_2024}. Modification in the application code (i.e., refactoring) also has an impact on execution time~\cite{traini_how_2021}. While most prior work aimed to reduce variability by eliminating its sources, our approach instead treats performance variability as an inherent system characteristic and focuses on modeling it through predictive performance models.

Multiple studies have attempted to achieve performance predictability in cloud and edge settings. Zhao et al.~\cite{zhao_cloud_2021} explored the variability introduced by OS and kernel changes, developing gradient-boosted models to predict performance changepoints in advance. Mohamed et al.~\cite{mohamed_end--end_2021} built machine learning models to predict microservice performance using a subset of monitoring metrics, though their selection was limited and arbitrarily chosen. Fu et al.~\cite{fu_progress-based_2019} estimated the remaining execution time based on application-reported progress, which required code modifications and did not support pre-submission predictions. Wang et al.~\cite{wang_predicting_2018} used CPU benchmarks to infer performance of unseen CPUs based on their specifications, but did not address other components such as GPUs. Krasniqi et al.~\cite{zheng_end--end_2024} applied neural networks to predict complex network performance, yet did not consider additional sources of variability such as CPU or memory contention. In contrast, our work focuses on building lightweight, accurate performance predictors based on carefully selected monitoring metrics. These predictors incorporate node specifications and account for multiple resource types affected by real-world co-location.

Other studies have used observability platforms to analyze performance metrics. Sukhija et al.~\cite{sukhija_event_2020} employed Kubernetes, Prometheus, and Grafana to detect incidents in real time using metrics such as CPU utilization. Mart et al.~\cite{mart_observability_2020} proposed a similar framework. Though effective for anomaly detection, these systems do not directly estimate task execution times. Our work complements these efforts by estimating the RTT of incoming application tasks based on live monitoring metrics, thus linking system state directly to application performance.

Efforts to improve scheduling decisions by integrating performance predictors have also emerged. Dauwe et al.~\cite{dauwe_hpc_2016} predicted application execution time and energy usage in multicore processors, accounting for interference among cores. However, their approach did not extend to network contention or other node-level components. Townend et al.~\cite{townend_invited_2019} developed hardware-based models to select optimal servers, with the aim of minimizing energy usage while maintaining performance, but did not explicitly predict the performance of the application. Sinan~\cite{zhang_sinan_2021}, a CPU-based autoscaler, used machine learning to dynamically allocate resources to meet QoS constraints. However, it only addressed CPU resources and did not provide pre-deployment performance predictions.

In summary, most existing approaches rely on a limited subset of monitoring metrics or require application-level instrumentation, such as exposing internal progress indicators. Few methods account for node-specific characteristics or the impact of co-location on performance, let alone address both sources of variability in combination. Our approach automatically analyzes the full set of monitoring metrics provided by the system to identify those most strongly associated with performance degradation. To our knowledge, no prior work applies this level of comprehensive analysis. The selected metrics are then used to train a variety of predictive models, resulting in a robust and automated framework for performance forecasting. Although our predictors are not yet integrated into scheduling mechanisms, they are designed to capture interference effects across multiple resource types (e.g., CPU and GPU) and adapt to dynamic co-location patterns. The resulting models are both lightweight and accurate, making them well-suited for real-time, performance-aware scheduling decisions.

\section{Methodology} \label{methodology}
\begin{figure*}
\centerline{\includegraphics[width=1\columnwidth]{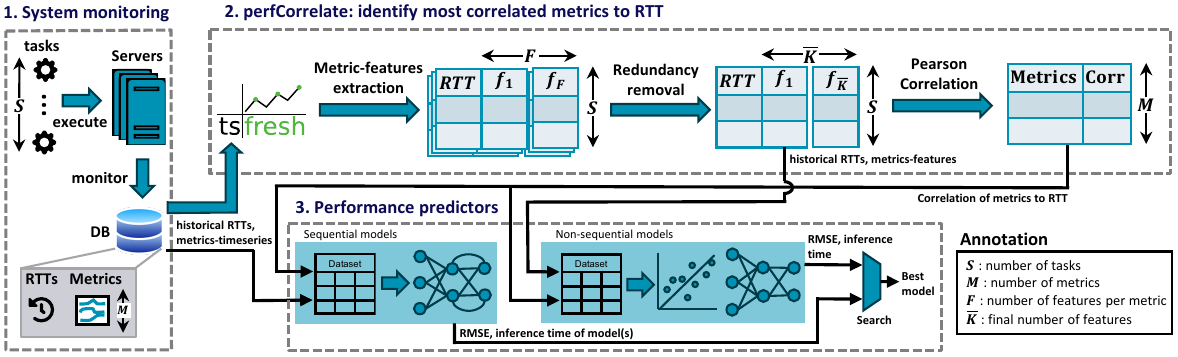}}
\caption{Overview of the proposed architecture for accurate performance prediction in edge computing environments. The system operates in three main stages: (1) \textbf{System monitoring:} The monitoring system continuously collects time-series data from each server, including metrics such as CPU, memory, network usage, and RTT values per task. This data is stored in a centralized database. (2) \textbf{Feature selection and correlation analysis (perfCorrelate):} From the complete set of monitoring metrics, time-series features are extracted. perfCorrelate computes the correlation between these features and historical RTTs to identify the most relevant metrics. Redundant features are removed to produce a compact and informative feature set. (3) \textbf{Predictor generation and evaluation:} Using the selected metrics, both sequential and non-sequential machine learning models are trained to predict task RTTs before execution. Sequential models use the raw time-series form of the metrics, while non-sequential models use the extracted features. All models are evaluated based on RMSE and inference time, and the best-performing one is selected for each application-node pair.} 
\label{fig:methodology}
\end{figure*}
%
Our work focuses on predicting the RTT for incoming tasks using historical monitoring metrics. In this context, RTT specifically refers to the time interval measured at the cluster gateway, which is a dedicated entry point within the cluster infrastructure responsible for receiving incoming network requests and routing them to appropriate server nodes. RTT is defined as the interval from the moment the request reaches the gateway ($t_{\text{start}}$) until the corresponding server response leaves the gateway ($t_{\text{end}}$). This measurement explicitly excludes any network transmission delays that occur between the client and the gateway. Any delays between the client and the cluster gateway are treated as external and fall outside the control of the cluster for optimization. Furthermore, in our current work, we assume that the variability of the input data does not significantly influence the RTT of the task. We consider it an external factor, independent of node heterogeneity or resource contention.

Figure~\ref{fig:methodology} shows the architecture of our methodology, highlighting the integration with existing cloud/edge computing components to ensure predictable performance. In the beginning, we are monitoring both the resource usage of the servers and the RTT of tasks towards applications. Afterwards, we use perfCorrelate~\cite{giannakopoulos_perfcorrelate_2025}, which analyzes all available monitoring metrics and extracts the most important ones for a specific application. In the end, we use the most important metrics to develop different performance predictors for each application and select the most efficient one (i.e. high accuracy, low overhead). 
%
\subsection{System monitoring}
Edge and cloud systems typically employ monitoring tools to track resource usage and identify potential failures in the infrastructure. These monitoring metrics, which capture the overall cluster state and are collected from various resource components, can help identify the sources of performance changes due to resource contention. Our monitoring system retrieves RTT and various resource metrics for each task, including CPU, RAM, GPU, and network usage. The resource usage metrics are recorded both before a task starts and throughout its execution. For CPU metrics, values from different cores are aggregated by calculating their mean. In this way, the general level of CPU load is calculated on each server. We capture the monitoring metrics for each task within the time interval $[t_{\text{start}} - t_{\text{offset}},~t_{\text{end}}]$, where \(t_{\text{offset}}\) represents the length of the historical state. The length of the historical state, measured in seconds, refers to the time interval preceding the task submission. The historical state of the monitoring metrics is essential for modeling system performance before submitting a task. Metrics that remain stable over time are discarded as they are deemed to be not correlated with performance. Our dataset comprises \(M\) monitoring metrics recorded for each of the \(S\) repetitions of tasks. 
\subsection{Feature selection and correlation analysis}
There are hundreds (i.e, 554) of metrics provided by the monitoring system, making their direct use in building machine learning models challenging due to increased dimensionality and computational complexity. To address this, we developed perfCorrelate~\cite{giannakopoulos_perfcorrelate_2025}, a tool that analyzes and computes the correlation of each monitoring metric with task performance. This approach identifies a subset of metrics that not only enhances the understanding of performance variability but also results in more efficient and accurate prediction models. perfCorrelate involves three main steps. In the first step, monitoring metrics are transformed from the time domain to the feature domain, enabling processing with conventional correlation algorithms such as Pearson correlation. In the second step, redundant metrics are removed, retaining only the relevant ones to accelerate further processing. In the last step, the correlation of each metric with task performance is calculated, identifying the most significant metrics.
\subsubsection{Metric-Features Extraction}
The resource metrics and RTTs for each task are retrieved from the database (i.e. monitoring system). Each of the \( M \) time series is converted into \( F \) statistical and temporal features, resulting in 785 feature-based attributes (e.g., mean and standard deviation) per time series. This transformation ensures that both the metric-related features and the RTT have the same dimensionality, allowing the use of conventional correlation methods. We employ tsfresh~\cite{christ_time_2018}, an open-source Python package, to automatically extract relevant attributes from time series data. Tsfresh offers a broad range of feature extraction methods to capture various characteristics of time series, such as trends, seasonality, and statistical properties. More details on the extracted features can be found in \cite{yang_anomaly_2021}. After this process, we obtain a set of extracted features for each monitoring metric per task.

\subsubsection{Redundancy Removal}
To reduce the high computational and memory costs of analyzing a dataset with 785 samples and 554 metric-related features, we apply two redundancy removal steps using Pearson correlation~\cite{benesty2009pearson}. Pearson correlation is a fast and simple method for checking relationships between metrics, which makes it useful for real-time analysis. However, it only detects linear patterns and may miss important non-linear relationships or interactions between metrics, possibly leading to overlooked information.

In the first substep, each of the \( M \) metrics is correlated with RTT separately. A correlation matrix is created for each metric using the \( F \) metric-related features and RTTs of tasks. Redundant features are eliminated by identifying pairs of features with a high inter-correlation greater than 90\%. In such case, the feature less correlated with RTT is removed. This process reduces the number of features per metric from \( F \) to \(\bar{F_i}\), where \( i \in [1, M] \) and \(\bar{F_i} \leq F\). The remaining metric-related features are then combined into a unified dataset consisting of \( S \) rows and \( K \) columns, where \( K = \sum_{i=1}^{M} \bar{F_i} \).

In the second substep, a correlation matrix is constructed that includes the metric-related features of all metrics and RTT. Features from different metrics that are strongly correlated with each other are removed, further reducing the dataset size. This results in a final dataset with \( S \) rows and \(\bar{K} \) columns, where \(\bar{K} \leq K \).

\subsubsection{Extract Correlations}
We apply Pearson correlation to the reduced unified dataset to correlate the metric-related features with RTT, enabling us to determine the significance of each monitoring metric. Instead of differentiating between positive and negative correlations, we use the absolute values of the correlation coefficients. Alternative correlation algorithms, such as Spearman's rank correlation, can also be utilized instead of Pearson correlation. For each monitoring metric, we calculate its correlation with RTT by selecting the maximum correlation value among its related features (those extracted by tsfresh). We then rank the metrics in descending order based on their correlation value, allowing us to identify the most significant metrics.
\subsection{Predictor generation and evaluation} \label{subsection:predictors}
We use the filtered monitoring metrics to build performance predictors through a design process that selects the most accurate models under strict latency constraints. Applications are treated as black boxes, meaning that we do not require access to their source code. This ensures that no changes are necessary to the applications, as the required data is gathered by standard monitoring systems that are already widely adopted in edge and cloud environments. To account for differences between applications, we train a separate model for each application type. Additionally, since each server has its own hardware, software, and set of co-located applications, performance can vary significantly from one node to another. As a result, we also develop separate predictors for each server. Ultimately, this leads to a dedicated performance predictor for each combination of application type and server node.

In order to identify the most suitable model per application/node pair, we iterate through different machine learning models, historical window lengths $t_{\text{offset}}$ and number of input features. Each model is evaluated based on its Root Mean Square Error (RMSE), defined as:
\begin{equation}
\text{RMSE}(Y, \hat{Y}) = \sqrt{\frac{1}{n}\sum_{i=1}^{n} (Y_i - \hat{Y}_i)^2}
\label{eq:rmse}
\end{equation}
%
where \( Y_i \) are the true values and \(\hat{Y}_i \) are the predicted values of RTT.
The models use as input \( X \), the most correlated monitoring metrics (i.e., top 10) identified by perfCorrelate, and predict \( y \), the RTT of new tasks.

The dataset is normalized, and any outliers are removed. Then it is divided into three parts: the training set, the validation set, and the test set. The training set is used to train the models, the validation set to select the best hyperparameters, and the test set to evaluate the models. We utilize keras-tuner~\cite{omalley2019kerastuner}, a hyperparameter optimization framework, to select the best hyperparameters for the performance predictors. Furthermore, we measure the inference time of each model, which refers to the time it takes for the model to generate a prediction after receiving input. The inference time is crucial for assessing whether the developed model can make predictions fast enough to be used in a real-time system.  

Our work distinguishes between two main categories of performance predictors: non-sequential models and sequential models. The division into non-sequential and sequential models is justified by the nature of the data: non-sequential models handle static relationships in structured data, while sequential models capture temporal patterns and dependencies present in time series data. This approach ensures flexibility in addressing both types of patterns for accurate performance prediction. Both categories predict the RTT of new tasks, but use the monitoring metrics in different forms. The input monitoring metrics are those identified by perfCorrelate. 

\subsubsection{Non-sequential Models}
This category includes models commonly used for analyzing structured or tabular data, where the order of the data is irrelevant. The selected models are: Linear Regression (LR)~\cite{linear_regression}, XGBoost (XGB)~\cite{xgboost}, Random Forest (RF)~\cite{ho1995random}, Support Vector Machine (SVM)~\cite{drucker1997support}, and Feedforward Neural Networks (FNN)~\cite{goodfellow2016deep}. These models were chosen to cover a diverse set of learning paradigms. LR offers a simple and interpretable baseline for capturing linear relationships. RF and XGB are tree-based ensemble methods known for their robustness to noise and ability to model complex, nonlinear interactions. SVM is particularly effective in high-dimensional feature spaces and is suitable for non-linear regression tasks, especially when the dataset is not very large. Lastly, FNNs are general-purpose, non-linear function approximators capable of learning intricate mappings between input monitoring metrics and RTT. The input features \( X \in (x_1, x_2, \ldots, x_M) \) for these non-sequential models consist of the extracted metric-features from tsfresh, where \( x_i \) represents the extracted feature of each monitoring metric during the historical state prior to task submission (i.e., the last 5 seconds).This dataset, initially generated by perfCorrelate to evaluate the importance of each monitoring metric, is reused for model training.
\subsubsection{Sequential Models}
%
This category includes models designed to work with sequential data, where the order of data points is important. The models we evaluate in this category are: Recurrent Neural Networks (RNN)~\cite{RNNs}, Long Short-Term Memory (LSTM) networks~\cite{hochreiter1997long}, Gated Recurrent Units (GRU)~\cite{wu2017introduction}, and Convolutional Neural Networks (CNN)~\cite{chung2014empirical}. These models are often used for time series forecasting, natural language processing, and other tasks that involve changes over time. We selected these models because they are good at learning patterns and dependencies in sequences. They can capture how values change over time and how past values affect future ones. This is useful in our case, where we want to understand how monitoring metrics evolve before a task is submitted. Each model takes as input a set of time series features \(\textbf{X} = (X_1, X_2, \ldots, X_M)\), where each \(X_i\) is the time series of a specific monitoring metric. These time series are collected over a fixed historical window before task submission (e.g., 5 seconds). All time series have the same length for all metrics and tasks, as they are collected over a fixed window (e.g., 5 seconds) before task submission.
\subsubsection{Model selection}
Each predictor is designed by exploring combinations of model types, features, and history lengths to minimize RMSE while meeting real-time inference constraints. The models are evaluated using two key metrics: RMSE and inference time. The objective is to select the model that offers the most accurate RTT predictions while maintaining a low inference latency. Inference time is particularly important as it directly contributes to the total RTT experienced by users. To prevent noticeable delays in scheduling, we require that the inference time does not exceed 1\% of the mean RTT of the application. This ensures that the prediction process remains lightweight and suitable for real-time environments.

The model selection process is formalized as an optimization problem:
\begin{equation}
\begin{aligned}
f_{\text{select}} = \underset{m, d, t_{\text{offset}}}{\mathrm{arg\,min}} \ \text{RMSE}(m(d, t_{\text{offset}})) \\
\text{subject to} \quad t_{\text{inference}}(m) \leq \mu_{\text{RTT}} \cdot \tau_{\text{inference}}
\end{aligned}
\label{eq:model_selection}
\end{equation}
where \(m\) denotes a machine learning model, \(d\) is the number of selected input features, and \(t_{\text{offset}}\) is the length of the historical window used. \(\text{RMSE}(m(d, t_{\text{offset}}))\) is the prediction error of the model. \(t_{\text{inference}}(m)\) is the model inference time and \(\mu_{\text{RTT}}\) is the mean RTT of the target application. The term \(\tau_{\text{inference}}\) sets the maximum allowed inference time as a fraction of the mean RTT.

%
\section{Experimental set up} \label{experiment_set_up}
In the upcoming section, we introduce the deployed applications, the installed frameworks, used infrastructure, and the workflow of our experiments.
\subsection{Applications}
In our experiments, we utilized five SPA applications: Upload, MotionCor2, FFT Mock, gCTF, and ctffind4. Each application gets a POST request with the provided input configuration, executes the specified algorithm, and returns the result. These applications are designed to handle one task at a time. The functionality of each application is the following:

\begin{itemize}
    \item \textbf{Upload:} Receives an image file from the client and stores it at the target worker node.
    \item \textbf{MotionCor2:} Addresses drift and beam-induced motion during SPA image acquisition, which can degrade image quality~\cite{BRILOT2012630}. Images are captured as movies of multiple frames, and MotionCor2 corrects the motion before combining them into a single integrated image~\cite{Zheng2017MotionCor2}. The GPU-based algorithm outputs a single-frame image stored as an MRC file~\cite{crowther_mrc_1996} for further processing. We use MotionCor2 v1.5.0.
    \item \textbf{FFT Mock:} Simulates lightweight image processing tasks used during SPA acquisition to maintain optimal microscope conditions. This mock version replicates the computational load of proprietary algorithms by duplicating and shifting pixels of a single MRC image, then calculating the induced shift using Fast Fourier Transform (FFT) and pixel-wise operations.
    \item \textbf{gCTF:} Corrects the Contrast Transfer Function (CTF) in EM image post-processing, determining optical state parameters to enhance image quality and ensure accurate downstream analysis. We used gCTF v1.18, a GPU-accelerated real-time CTF determination and correction algorithm~\cite{zhang_gctf_2016}.
    \item \textbf{ctffind4:} Another application for obtaining CTF information from an MRC image, ctffind v4.1.14~\cite{ROHOU2015216}, operates exclusively on the CPU.
\end{itemize}

The image data originates from a vitrified EM grid containing a protein sample, acquired using a Glacios electron microscope equipped with a Falcon4i detector and EPU software~\cite{deng_ispa_2021}. The experiments were conducted offline in a separate compute environment from the microscope. All applications use input files in the MRC format, a standard for storing electron microscopy image and volume data~\cite{crowther_mrc_1996}. The fraction files are used as input for the MotionCor2 and Upload applications, while the single integrated MRC image served as input for the other applications. The fraction files are 897MB each, and the integrated files are 65MB. Our dataset comprised 98 images.
\subsection{Frameworks}
Kubernetes~\cite{Kubernetes} is an open-source framework designed to orchestrate distributed computational resources, known as nodes, and facilitate application deployment using pods. Each pod can contain one or more containers, leveraging container technologies such as Docker~\cite{Docker} or containerd~\cite{Containerd}. In a Kubernetes cluster, the master node oversees the entire system, managing the allocation of pods to resources, while the worker nodes host the assigned pods. GPU resource scheduling is supported through the installation of GPU device plugins provided by GPU vendors.

Prometheus~\cite{Prometheus} is an open-source monitoring system that integrates seamlessly with Kubernetes to collect time-series data from both applications and infrastructure components. The Prometheus architecture includes a server and various monitoring targets. The server periodically scrapes metrics from these targets over HTTP in a text-based format and stores the data in a local time-series database. Users can access this collected data via the built-in web UI, dashboard tools like Grafana~\cite{Grafana}, or directly through the Prometheus HTTP API. Prometheus provides a comprehensive view of the cluster state by monitoring nodes and pods for CPU, RAM, disk, and network usage. Additionally, GPU metrics are gathered using the NVIDIA Data Center GPU Manager (DCGM) exporter~\cite{dgcm}. In our setup, monitoring metrics are recorded at a frequency of one measurement per 200 milliseconds. 

The overhead of the monitoring system comprises the resource consumption of both the Prometheus server and the monitoring targets. In our setup, the Prometheus server consumes about 150\% CPU (1.5 CPU cores) and 15GB of RAM. The monitoring targets, including node and DCGM exporters, collectively use approximately 35\% CPU and 300MB of RAM. The network bandwidth for sending metrics from each node to the Prometheus server is approximately 115KBps.
\subsection{Compute infrastructure}
\begin{table*}
\caption{Specifications of infrastructure components.}
\begin{center}
\centerline{\begin{tabular}{|l|l|l|c|c|c|l|l|}
\hline
\textbf{Name} & \textbf{Type} & \textbf{Model/Processor} & \textbf{Cores} & \textbf{RAM} & \textbf{Disk} & \textbf{GPU} \\
\hline
Switch & Physical & Dell PowerConnect 5448 & - & - & - & - \\
\hline
Routing & Virtual & Intel Xeon X3430  & 3 & 4 GB & HDD  & -  \\
\hline
Master & Virtual & Intel Xeon Silver 4109T & 8 & 32 GB & SSD &  - \\
\hline
Monitor & Virtual & Intel Xeon Silver 4109T & 8 & 78 GB & SSD &  - \\
\hline
Worker-1 & Physical  & Intel Core i7-7700 & 4 & 16 GB & HDD & Tesla K20c \\
\hline
Worker-2 & Physical & 2xIntel Xeon E5-2637 & 4 & 24 GB & HDD & Tesla M40 \\
\hline
Worker-3 & Physical & Intel Core i9-14900K & 32 & 128 GB & SSD & GeForce RTX4090 \\
\hline
Worker-4 & Virtual & Intel Core i5-9600 & 4 & 12 GB & SSD & - \\
\hline
Worker-5 & Physical & 2xIntel Xeon E5504 & 8 & 24 GB & HDD & - \\
\hline
Worker-6 & Virtual & Intel Xeon Silver 4209T & 8 & 24 GB & SSD & - \\
\hline
Worker-7 & Virtual & Intel Xeon Gold 6138T & 16 & 64 GB & SSD & - \\
\hline
Worker-8 & Virtual & Intel Xeon Silver 4209T & 8 & 24 GB & SSD & - \\
\hline
Worker-9-29 & Physical & Intel Xeon X3430 & 4 & 4/8 GB & HDD & - \\
\hline
\end{tabular}}
\label{tab:specifications}
\end{center}
\end{table*}

\begin{figure}
\centerline{\includegraphics[width=0.5\columnwidth]{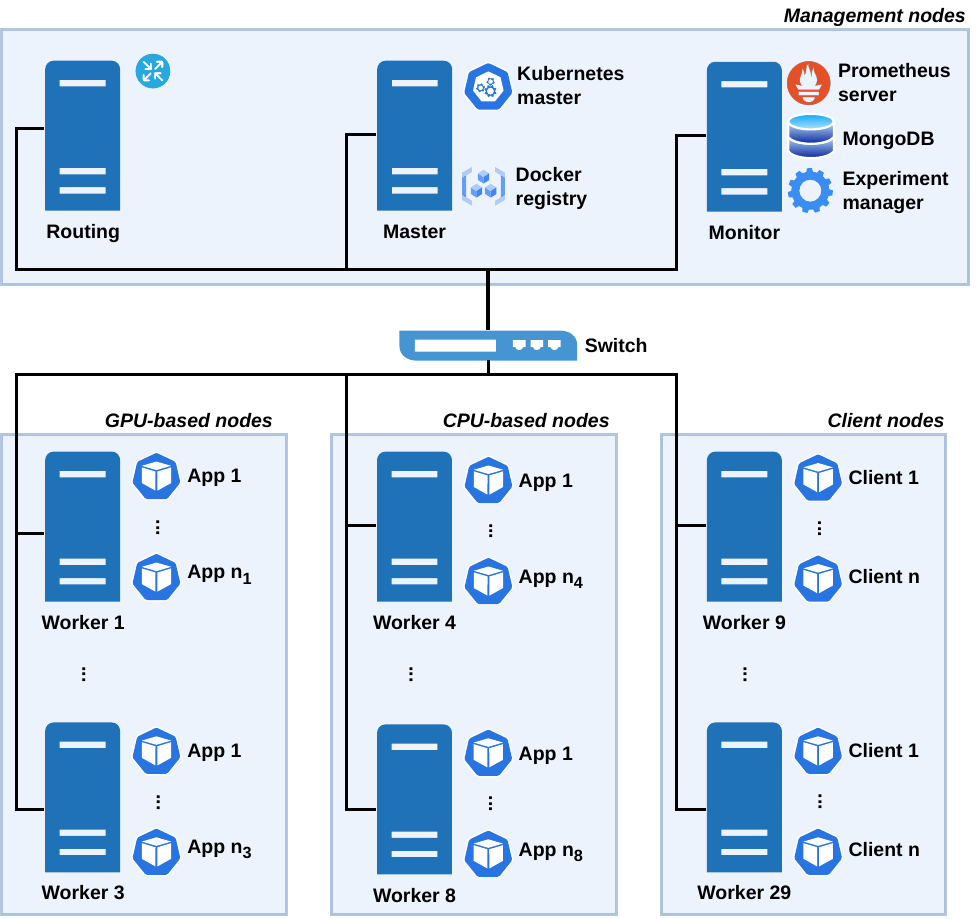}}
\caption{Structure of the used infrastructure. The servers are connected through 1 Gbps Ethernet link.}  
\label{fig:cluster_structure_dc}
\end{figure}

The structure of the set-up used for the experiments is depicted in figure \ref{fig:cluster_structure_dc} and the detailed specifications of the components can be found in Table~\ref{tab:specifications}. The set-up includes 32 servers, both physical and virtual, which are connected to 1 Gbps links. The Kubernetes cluster has one master and 30 worker nodes. Note that the size of the cluster does not significantly impact this performance predictability study, as the predictors are designed and trained specifically for each individual node. In larger-scale deployments, this design enables scalability by avoiding centralized bottlenecks, since each node can host and update its own predictor independently. The set-up is organized as follows: 

\begin{itemize}
    \item \textbf{Routing}: A virtual machine that manages internal network routing. This server is not part of the Kubernetes cluster.
    \item \textbf{Master}: This virtual machine serves as the master node of the Kubernetes cluster and hosts the local Docker registry, crucial for distributing custom Docker images within the cluster.
    \item \textbf{Monitor}:A virtual machine within the Kubernetes cluster that hosts the Prometheus server, the Experiment Manager, and a MongoDB database for storing experiment data. The Experiment Manager deploys and terminates the pods related to our experiments.
    \item \textbf{Worker-*}: A mix of physical and virtual worker nodes responsible for deploying applications and generating tasks. These nodes also run Prometheus agents to monitor resource usage and export data to the Prometheus server. Workers 1-8 are dedicated to hosting applications, while Workers 9-29 host the application clients. Workers 1-3 are equipped with GPUs, allowing them to support GPU-based applications.
\end{itemize}

\subsection{Experiment workflow}
In our experimental setup, we deploy as many instances of each application as possible across Workers 1-8, continuing until the available resources on each node are fully utilized. Resource allocation for these applications is unrestricted, meaning that applications consume minimal resources while idle but can utilize all available resources during active execution. Additionally, dedicated client pods remain inactive by default and initiate tasks only upon receiving explicit commands. To systematically explore performance under varying resource contention, we gradually increase the number of co-located applications on each node until its resources are completely exhausted. Subsequently, we gradually reduce contention by sequentially stopping task generation for individual applications. To clearly structure this process, the experiments are divided into distinct workload stages.

A \textit{workload stage} is defined as a specific period during which a subset of deployed applications remains actively processing tasks. Each application instance is associated with a dedicated client pod, responsible for initiating requests, waiting for responses, and subsequently pausing for a randomized duration, $t_{\text{wait}}$, before submitting the next task. The waiting interval $t_{\text{wait}}$ is randomly selected from the interval $[0,t_{\text{max}}]$, with the maximum waiting time, $t_{\text{max}}$, set individually per application as follows: 40 seconds for \textit{upload}, 6 seconds for \textit{ctffind4}, 20 seconds for \textit{FFT Mock}, 10 seconds for \textit{gCTF}, and 10 seconds for \textit{MotionCor2}. These values were empirically determined based on preliminary experiments to ensure that applications generate task loads frequently enough to compete for shared resources. This competition induces measurable performance interference among the co-located applications, which is necessary for observing realistic contention patterns and evaluating the effectiveness of our performance prediction.  The RTT for each task is recorded on the client side and stored in a MongoDB database. Although RTT, as previously defined in Section~\ref{methodology}, can also be inferred on the server side using tracing tools such as Jaeger~\cite{jaegertracing}, implementing this capability is planned for future research.

Our Experiment Manager is deployed as a pod and controls the experiment workflow. It starts by deploying the application and client pods on each node using the Kubernetes API. Each node is then evaluated by creating a distinct process that activates the client(s) for the first workload stage. This process periodically checks (every 30 minutes) if enough samples have been collected through CONFIRM~\cite{maricq_taming_nodate}. CONFIRM calculates the minimum number of task repetitions required to reach a confidence level of \(\alpha\)\% (e.g., 95\%), ensuring that the observed median does not deviate by more than \(r\)\% (e.g., 5\%) from the true median of the RTT distribution. When sufficient samples are collected for the current stage, the Experiment Manager activates the client(s) for the next stage, gradually increasing the number of active applications and resource consumption. Once enough samples have been collected during the final workload stage, which corresponds to the last phase of the experiment, the Experiment Manager removes all client and application pods to conclude the execution.

At the end of the experiment, our methodology is applied to the collected data as described in Section~\ref{methodology}. RTT data and monitoring metrics are retrieved from MongoDB and Prometheus, respectively. Different datasets per application are generated by matching the monitoring metric state with RTTs for windows $t_{\text{offset}}$ equal to 5, 20, and 60 seconds. The selection of these windows is arbitrary. The correlation of each monitoring metric is calculated by perfCorrelate. Afterwards, different models are trained for each window and number of input features to identify the best one. Depending on the model type, the dataset takes different forms (see Section~\ref{subsection:predictors}). In each case, the dataset is normalized (between 0 and 1) by using MinMax normalization, and outliers are removed (z-score $>$ 3). After removing outliers, the sample counts for each application are as follows: Upload (515), gCTF (3,123), MotionCor2 (3,294), ctffind4 (5,968), and FTT Mock (776). The variation in sample numbers is due to the required repetitions specified by CONFIRM to accurately capture the performance variability of each application. The dataset is divided into three parts: 80\% for the training set, 10\% for the validation set, and 10\% for the test set. The selected model is the one that achieves the lowest RMSE while keeping its inference time below 1\% of the mean RTT.

\section{Results} \label{results}
In this section, we present the accuracy and overhead in terms of time of various machine learning models used to predict the RTT of tasks. We also examine how the number of input monitoring metrics and the length of historical windows influence the accuracy of the models. Finally, we demonstrate how these models generalize across different servers and adapt to dynamic co-location scenarios.
\subsection{Building predictors with non-sequential models}
\begin{figure*}
\centerline{\includegraphics[width=0.8\columnwidth]{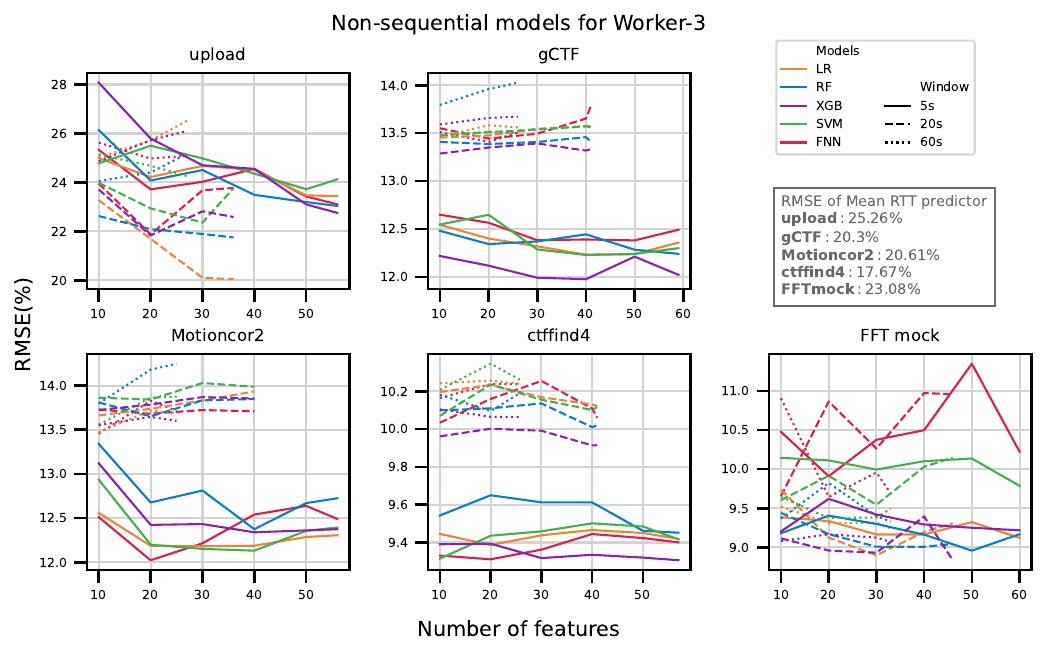}}
\caption{RMSE of non-sequential models for each application on Worker-3, showing how increasing the number of input features and history window length affects prediction accuracy. The plot also illustrates the RMSE when the mean RTT is used as the prediction.}  
\label{fig:classic_models}
\end{figure*}
We evaluate the accuracy of several non-sequential models across different combinations of input features and historical window lengths. Figure~\ref{fig:classic_models} presents the RMSE values for each model as the number of input features (i.e., monitoring metrics) gradually increases. The evaluation is performed for each application running on Worker-3. The models used include linear regression (LR), random forest (RF), XGBoost (XGB), support vector machine (SVM), and feed-forward neural network (FNN). Each model is initially trained using the ten most correlated features to RTT variability, and then incrementally re-trained from scratch by adding the next ten most correlated features until all available features are included. Note that the input features are sorted based on the correlation using perfCorrelate. For each configuration, we evaluate the performance of the model using three historical window lengths: \(t_{\text{offset}} = 5\), 20, and 60 seconds before task submission. The figure also includes the RMSE achieved when no monitoring metrics are used, where the model always predicts the mean RTT for each application.

Our results reveal that the best performing configuration (model, number of features, window), defined by the lowest RMSE, varies depending on the application. For example, XGBoost with a 5-second window yields the lowest error for gCTF, while linear regression with a 20-second window performs best for the upload application. This highlights the importance of tailoring the predictor to each application. XGBoost achieves the highest overall prediction accuracy, approximately 90\%, for ctffind4, using a 5-second window and all relevant monitoring metrics. The total number of features differs between historical windows because perfCorrelate filters out metrics that are not correlated with RTT in each specific window. As a result, the number of features used in the training depends on the chosen window size.

We observe that for some applications, such as ctffind4, gCTF, and MotionCor2, the accuracy of the model remains relatively constant as more features are added. In contrast, for applications like FFT Mock and upload, RMSE varies significantly, indicating a higher sensitivity to selected features. In some cases, adding more metrics leads to poorer performance because the additional features introduce noise that reduces the accuracy of the model. Therefore, feature addition should be stopped once extra features do not lead to a clear improvement or even cause a drop in RMSE. This strategy helps to keep the model lightweight and fast by reducing its complexity.

These findings highlight the need to explore and optimize the model type, feature set, and historical window per application and node to achieve the best accuracy. Automating this selection process is essential for a scalable deployment. Finally, we observe that models using monitoring metrics significantly outperform the baseline mean prediction approach, whose RMSE is directly influenced by the variability of the RTT distribution.
\subsection{Building predictors with sequential models}
\begin{figure*}
\centerline{\includegraphics[width=0.8\columnwidth]{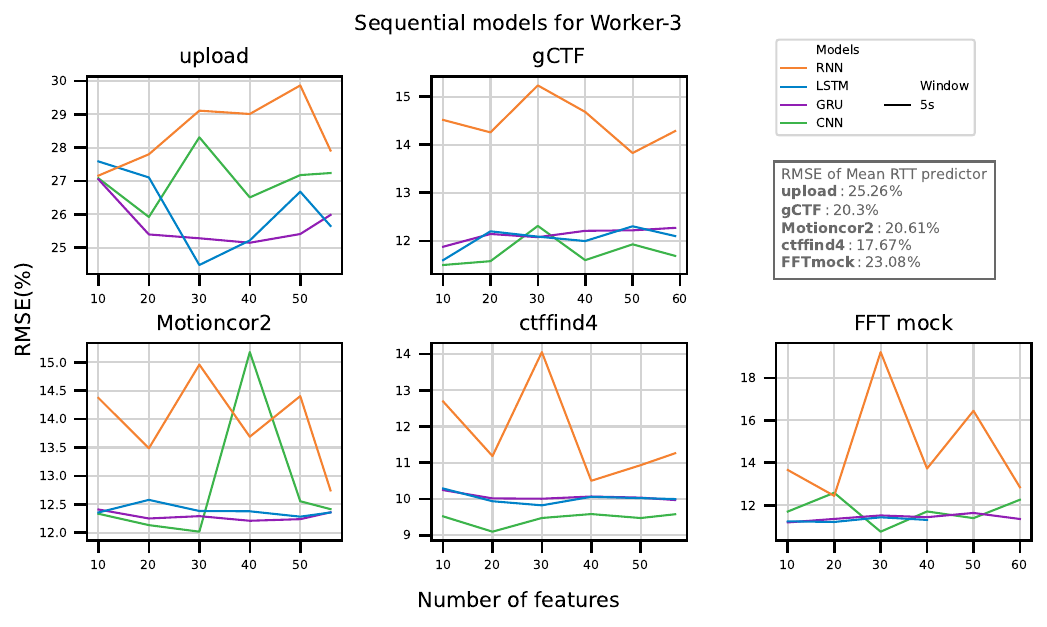}}
\caption{RMSE of sequential models for each application on Worker-3, showing how increasing the number of input features and history window length affects prediction accuracy. The plot also illustrates the RMSE when the mean RTT is used as the prediction.} 
\label{fig:sequential_models}
\end{figure*}
Non-sequential models use statistical and temporal features extracted from monitoring metrics, rather than processing the raw time series data directly. This transformation from the time domain to the feature domain can result in information loss, potentially increasing prediction error. To address this limitation, we develop four sequential models: RNN, LSTM, GRU, and CNN. These models use monitoring metrics as time series input, allowing them to learn more complex patterns between resource usage and performance variability.

Figure~\ref{fig:sequential_models} shows the RMSE of the sequential models for each application running on Worker-3, using a 5-second history window before task submission and varying numbers of input features. Although we also considered longer windows of 20 and 60 seconds, these were excluded due to their high computational cost and extended training time.

For most applications, the sequential models achieve similar or slightly better accuracy compared to the non-sequential ones. This suggests that, even though non-sequential models rely on features such as the mean instead of raw time-series data, the transformation does not necessarily result in significant information loss. However, we observe that RMSE values for sequential models tend to fluctuate more as additional features are added, a pattern that is less common in non-sequential models. This behavior is likely due to the fact that sequential models generally require larger datasets to achieve stable and accurate results. Evaluating the accuracy of these models on larger datasets is left for future work. For the gCTF application, the CNN model achieves a lower RMSE than all non-sequential models. This shows that it is important to evaluate both sequential and non-sequential models in order to find the most accurate predictor for each application.

Sequential models are capable of capturing more detailed and complex relationships between monitoring metrics and task performance, but they also require more data and computational resources. In scenarios where data is limited or computational efficiency is a priority, non-sequential models may offer a more practical solution. Therefore, the choice between model types should be guided by both the available dataset size and the constraints of the deployment environment.
\subsection{Training and inferences of models}
Although accuracy is essential for selecting the right performance predictor, its training and inference times determine whether it can be used in real-time systems. In real-time pipelines, such as task scheduling, model inference must be fast enough to avoid introducing delays. In our methodology, we require the inference time to remain under 1\% of the total RTT. Training time is also important, especially when new data must be included to maintain prediction accuracy. Inference time is influenced by model complexity, while training time depends on model type, dataset size, and the range of hyperparameters. Both are also affected by system characteristics, such as processor speed and whether a GPU is used.

\begin{figure}
\centerline{\includegraphics[width=0.5\columnwidth]{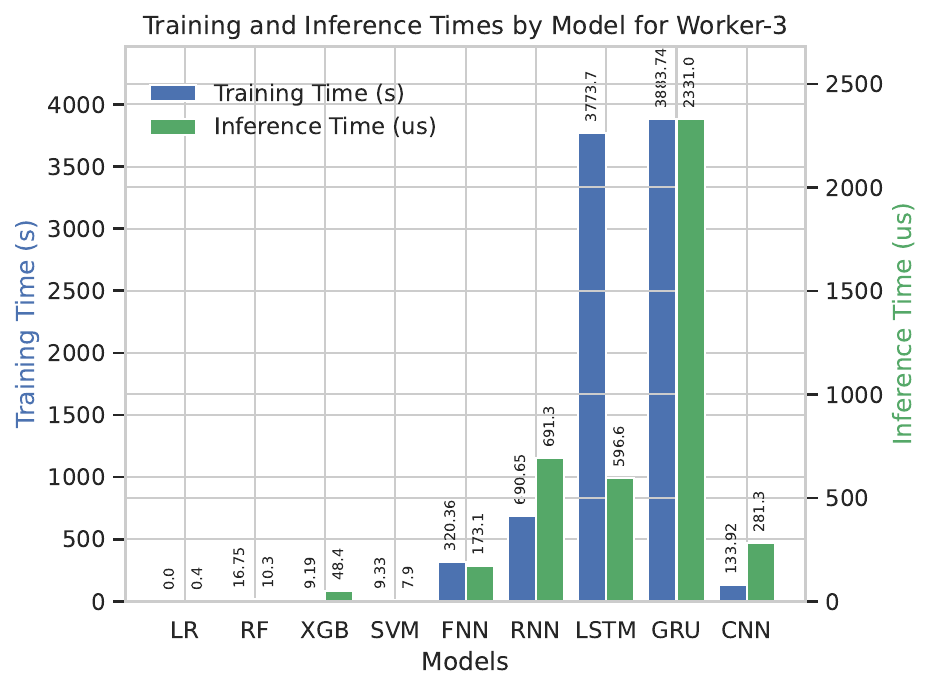}}
\caption{Training and inference times of sequential and non-sequential models for MotionCor2 on Worker-3.}  
\label{fig:overhead}
\end{figure}

Figure~\ref{fig:overhead} shows the training and inference times of various models for the MotionCor2 application on Worker 3, using an arbitrary window and number of monitoring metrics (5-second window and 10 metrics). The models were trained on a machine equipped with an Intel Core i9-14900K processor and an Nvidia GeForce RTX 4090 GPU. The x-axis lists the different models, the left y-axis shows the training time in seconds, and the right y-axis shows the inference time in microseconds. Models such as LR, RF, XGB, and SVM were trained on the CPU, while FNN and the sequential models (RNN, LSTM, GRU, CNN) were trained on the GPU.

For MotionCor2, which has an average RTT of 1.7 seconds in Worker 3, the FNN model adds only 173 microseconds of inference time, which is less than 0.01\% time overhead. Even for ctffind4, the application with the shortest RTT at around 636 milliseconds, FNN still introduces minimal delay. Sequential models, particularly GRU, have longer inference times due to their higher complexity, but their overhead remains low, staying below 0.4\% of the task RTT. This demonstrates that both sequential and non-sequential models can make predictions in real-time pipelines, where tasks operate on the scale of seconds or milliseconds. These predictors can estimate task RTTs with high accuracy and minimal inference time.

Training time becomes more important when the system needs to respond to new data, such as during the deployment of new applications or changes in workload. In such cases, non-sequential models are generally faster to train. For example, XGBoost completes the training in about 50 seconds, while GRU can take up to one hour. Since models must be retrained to reflect evolving performance conditions, training times can increase as the dataset grows. To manage this, training can be scheduled during low-load periods in the cluster. Alternatively, online learning methods can be applied, especially for neural networks, to update the model incrementally using only the latest samples.

It is important to note that the training times presented correspond to a single model instance. During hyperparameter tuning, multiple models must be trained and evaluated, which significantly increases the overall training time. To keep pace with changing co-location patterns on each node, training must be fast enough to allow timely model updates. In our setup, where co-location changed every few hours, we excluded models such as LSTM and GRU from real-time retraining due to their long training times. Currently, our methodology is applied offline, after the data collection phase is completed. However, in a production environment, data collection and model training need to run in parallel so that predictions can be continuously updated and available throughout system operation. If training takes too long, the co-location scenario may have already changed by the time the model becomes usable, requiring retraining to maintain accuracy. This underscores the importance of balancing model complexity and accuracy with training efficiency, especially in dynamic environments where system conditions evolve rapidly.
\subsection{Generalization of models among nodes}
\begin{table*}[]
\caption{Best combination of model ($m$), number of monitoring metrics ($d$) and historical window ($t_{\text{offset}}$) among worker nodes.}
\begin{center}
\scriptsize
\begin{tabular}{|c|ccc|ccc|ccc|ccc|ccc|}
\hline
& \multicolumn{3}{c|}{upload} & \multicolumn{3}{c|}{gCTF} & \multicolumn{3}{c|}{Motioncor2} & \multicolumn{3}{c|}{ctffind4} & \multicolumn{3}{c|}{FFT mock} \\ \cline{2-16}
Node & \multicolumn{1}{c|}{$m$} & \multicolumn{1}{c|}{$d$} & \multicolumn{1}{c|}{$t_{\text{offset}}$} & \multicolumn{1}{c|}{$m$} & \multicolumn{1}{c|}{$d$} & \multicolumn{1}{c|}{$t_{\text{offset}}$} & \multicolumn{1}{c|}{$m$} & \multicolumn{1}{c|}{$d$} & \multicolumn{1}{c|}{$t_{\text{offset}}$} & \multicolumn{1}{c|}{$m$} & \multicolumn{1}{c|}{$d$} & \multicolumn{1}{c|}{$t_{\text{offset}}$} & \multicolumn{1}{c|}{$m$} & \multicolumn{1}{c|}{$d$} & \multicolumn{1}{c|}{$t_{\text{offset}}$} \\ \hline
\multicolumn{1}{|l|}{Worker-1} & RF & 50 & 60s& CNN & 40 & 5s& FNN & 60 & 5s& FNN & 70 & 5s& LR & 60 & 5s\\ \hline
\multicolumn{1}{|l|}{Worker-2} & XGB & 50 & 20s& RF & 50 & 5s& RF & 59 & 5s& XGB & 65 & 5s& CNN & 10 & 5s\\ \hline
\multicolumn{1}{|l|}{Worker-3} & LR & 36 & 20s& CNN & 10 & 5s& CNN & 30 & 5s& CNN & 20 & 5s& XGB & 46 & 20s\\ \hline
\multicolumn{1}{|l|}{Worker-4} & RF & 30 & 5s& - & - & -& - & - & -& RF & 30 & 5s& RF & 20 & 5s\\ \hline
\multicolumn{1}{|l|}{Worker-5} & RNN & 46 & 5s& - & - & -& - & - & -& CNN & 40 & 5s& XGB & 30 & 5s\\ \hline
\multicolumn{1}{|l|}{Worker-6} & RF & 10 & 5s& - & - & -& - & - & -& XGB & 30 & 5s& XGB & 37 & 60s\\ \hline
\multicolumn{1}{|l|}{Worker-7} & CNN & 30 & 5s& - & - & -& - & - & -& CNN & 20 & 5s& RF & 20 & 5s\\ \hline
\multicolumn{1}{|l|}{Worker-8} & RF & 20 & 5s& - & - & -& - & - & -& XGB & 10 & 5s& XGB & 10 & 60s\\ \hline
\end{tabular}
\label{tab:node_generalization}
\end{center}
\end{table*}
In the previous sections, we analyzed the RMSE of different models for applications running on Worker-3. Table~\ref{tab:node_generalization} extends this analysis by evaluating how well the best-performing models generalize across different worker nodes. It shows the optimal combination of model type, number of input features, and historical window for each application on each node. The selected model in each case is the one with the lowest RMSE while keeping inference time below 1 percent of the mean RTT.

Applications such as gCTF and MotionCor2 are only deployed on Worker-1 through Worker-3, as they require GPU support. From the results, it is clear that no single model configuration consistently achieves the best accuracy across all nodes. These differences are mainly caused by the unique co-location scenarios on each node, as well as node heterogeneity, which lead to different performance behaviors.

The mean RTT of the same application can vary significantly between nodes due to different hardware capabilities. As a result, the best history window length for prediction also varies, which directly impacts the accuracy of the model. These findings confirm that model performance is sensitive to the specific environment of each node, including both hardware and software configurations. Our methodology takes this into account by building a separate model for each application-node pair. In theory, including node configuration as additional input features could allow the use of a shared model across nodes. However, in our case, the number of heterogeneous nodes is too small to provide enough data to train such a model effectively. 

The predictors can be hosted on a central node or deployed separately on each worker node. Although central hosting may seem simpler, it does not scale efficiently. As the number of worker nodes grows, the computational cost of training increases, and the network overhead from sending monitoring metrics to a central location becomes significant. In contrast, a distributed setup allows each node to host and manage its own models. Even though hardware capabilities vary across nodes, with some having GPUs and others not, this approach remains flexible. On nodes with limited resources, simpler models like XGBoost can be used, as they still provide high accuracy while avoiding the long training times required by complex neural networks on CPU-only systems.

The main benefit of the distributed approach is that monitoring metrics remain local to each node, reducing network traffic. Only the prediction results, which are much smaller in size, need to be sent to a centralized scheduler for decision-making. This improves scalability and ensures that the computational workload of model training is shared across the cluster, rather than concentrated on a single machine.
\subsection{Dynamic co-location change and prediction accuracy}
\begin{figure*}
\centerline{\includegraphics[width=0.8\columnwidth]{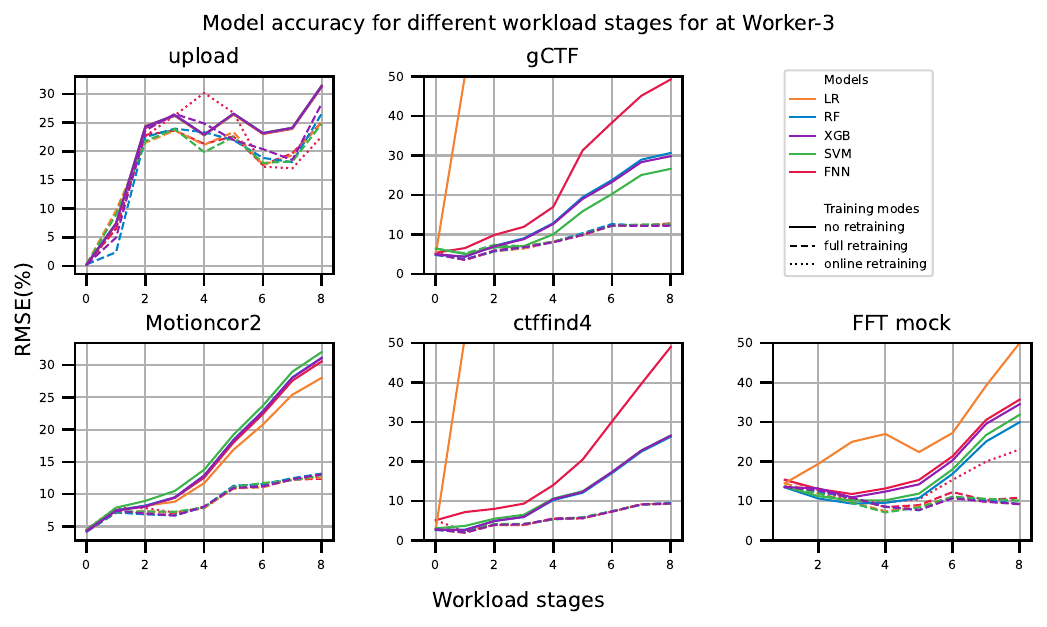}}
\caption{Accuracy of non-sequential models for different training modes as new co-locations occur on Worker-3}  
\label{fig:colocation}
\end{figure*}
The co-location of applications on a worker node can change over time as new applications are deployed or existing ones are removed. In addition, workload patterns, such as the rate of incoming tasks, can significantly affect the resource usage of running applications. These changes can cause performance variability, making it important to keep performance predictors up to date. To emulate this dynamic behavior, we gradually activate additional application instances on the worker nodes. We refer to the period during which a specific set of applications is active as a \textit{workload stage}.

Given their lower overhead and similar accuracy compared to sequential models, we focus this analysis on non-sequential models. We evaluate the accuracy of the model in three training modes: \textit{no retraining}, \textit{full retraining}, and \textit{online training}. In the no retraining mode, the model is trained once during the first workload stage and reused in all subsequent stages without updates. In full retraining, the model is updated whenever a new workload stage occurs. This involves re-training the model from scratch using all available data collected, including tuning hyperparameters (such as the number of layers in a feedforward network). Online training, used only for neural networks, updates model weights using only the new data collected since the last update, without revisiting the full dataset.

Figure~\ref{fig:colocation} shows the RMSE of different models as the workload on Worker-3 evolves through multiple stages. The x-axis represents the activation of new application instances over time, while the y-axis shows the prediction error. Each line corresponds to one of the three training modes. We observe that when models are not updated, the prediction error increases sharply after each new workload stage. This is due to the limited training data available from earlier stages, which does not reflect the new resource contention patterns. In some cases, such as \texttt{gCTF} and \texttt{ctffind4}, linear regression fails completely after the first stage, highlighting its inability to generalize without retraining.

In contrast, models that are updated during each stage, either through full retraining or online training, show a much slower increase in error. This demonstrates their ability to adapt to new co-location patterns and maintain higher accuracy. Full retraining generally results in slightly lower RMSE than online training, which is expected since it rebuilds the model with the full dataset and re-optimizes the hyperparameters based on the new data. However, even with continuous updates, RMSE increases as the workload stages progress due to rising resource contention from co-located applications. This contention leads to greater variability of the RTT, where similar monitoring metric patterns can produce significantly different outcomes, making the input-output relationship more difficult to learn~\cite{maricq_taming_nodate}. From a machine learning perspective, this variability corresponds to label noise or concept drift~\cite{Lu2018LearningUC}, which reduces the ability of the model to generalize and increases prediction error.

The main advantage of online training is that it updates the model using only recent data, keeping training time relatively stable as the system runs. Full retraining, on the contrary, becomes slower over time, since it uses the entire dataset, which grows as more workload stages pass. In real-time systems where monitoring data accumulates quickly, online training offers a more scalable solution. For models that do not support online training, an alternative is to manage dataset size by removing redundant entries and adding new samples only when co-location scenarios change.

In summary, to maintain accurate performance predictions in dynamic environments, models must be regularly updated. Online training provides an efficient way to achieve this with low overhead when supported. For other models, periodic full retraining remains necessary to keep up with changes in workload and resource sharing.
\subsection{Predict remaining RTT}
\begin{figure}
\centerline{\includegraphics[width=0.5\columnwidth]{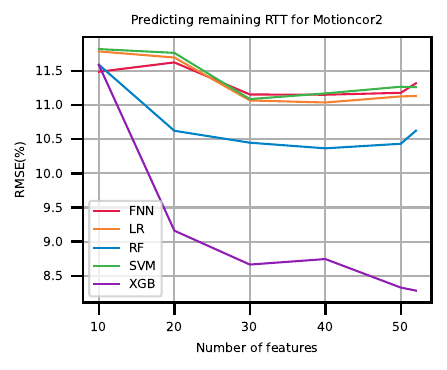}}
\caption{Accuracy of models for prediction the remaining RTT of MotionCor2 deployed on Worker-3}  
\label{fig:remaining_time}
\end{figure}
In the final part of our study, we develop non-sequential models to predict the remaining RTT of tasks that are already running. These predictions are useful for refining earlier RTT estimates made before task execution. We use both monitoring data from the historical state before task submission and data collected during the actual execution of each task. For this analysis, we assume that each task has already completed halfway. We extract monitoring metrics from 5 seconds before task submission until the time the task reaches 50 percent completion. The methodology used to select the number of features and machine learning model remains the same as in Section~\ref{methodology}. To estimate when a task reaches 50 percent of its total run-time, we rely on previously predicted RTT values based on historical metrics. If no predictions are available, this midpoint can be estimated using the average RTT of previous tasks.

Figure~\ref{fig:remaining_time} shows the RMSE of different models that predict the remaining RTT of the \texttt{MotionCor2} application on Worker-3. The x-axis shows the number of input features, while the y-axis shows the RMSE for each model. Among the trained models, XGBoost achieves the best result, with an RMSE of approximately 8.2\% when using all relevant monitoring metrics selected by perfCorrelate. One key insight from this figure is that the models that predict the remaining RTT have a lower error (ranging from 12\% to 8.2\%) than those that predict RTT before task submission. This improvement is justified because the models have access to more up-to-date information. Another important observation is that while CNN was the most accurate model for the prediction of RTT before task submission (see Figure~\ref{fig:classic_models}), XGBoost performs better when predicting the remaining RTT. This highlights that the best model can vary depending on the type of prediction, even for the same application and node.

These results highlight the importance of updating the predictions as the task progresses. They also show that different models may perform better at different stages. Therefore, using the same model for both pre-submission and mid-execution predictions may not give the best results, and it is important to evaluate multiple models for each case.
%
\section{Conclusions and future work} \label{conclusions}
In this study, we focused on building accurate performance predictors for edge computing applications, addressing the challenges posed by resource heterogeneity and dynamic co-location. Specifically, we examined the performance predictability of Single Particle Analysis (SPA) applications operating at near-real-time timescales (e.g., subsecond) and requiring diverse computational resources such as CPU and GPU. In edge environments, application performance is often unpredictable due to interference from co-located tasks and node heterogeneity. To tackle this, we proposed a methodology that automatically generates and selects predictors through a constrained optimization process that ensures both accuracy and real-time inference for each application-node pair.

The generated models are divided into two categories: sequential and non-sequential. These models are designed to predict the Round Trip Time (RTT) of tasks before submission, using historical monitoring data from the most relevant performance-related metrics, identified by perfCorrelate. In addition, we extended our analysis to predict the remaining RTT of tasks that are already in progress, using monitoring data collected during their execution. To ensure practical use in real-time scheduling, we identified the optimal configuration for each predictor by balancing accuracy and inference time. This includes selecting the appropriate model type, number of features, and history window length. Given the heterogeneous nature of the nodes in edge environments, we showed that different performance models are needed for each node and application. We also demonstrated the importance of model adaptability by comparing three training strategies: no retraining, full retraining, and online training.

Although this study focuses on SPA applications, the methodology is application-agnostic and can be applied to other domains. In future work, we plan to evaluate its generalizability across diverse edge workloads and explore more advanced machine learning models. We also aim to transition our approach to a fully run-time system that continuously trains and updates predictors as new data becomes available. This will allow us to study prediction accuracy and system overhead in greater detail, both in terms of resource usage and training or inference time. Ultimately, our goal is to integrate these predictive models into the Kubernetes scheduler, enabling more informed task placement decisions and improving overall system efficiency.
\section*{Acknowledgments}
This publication is part of the project ADAPTOR: Autonomous Distribution Architecture on Progressing Topologies and Optimization of Resources (with project number 18651 of the research Open Technology Programme which is (partly) financed by the Dutch Research Council (NWO). The image data used for the experiments was acquired at Thermo Fisher Scientific (Eindhoven, The Netherlands). The computational experiments, including execution of the described algorithms, were executed at the Eindhoven University of Technology (Eindhoven, The Netherlands).

\bibliographystyle{unsrt}  
\bibliography{ref}

\end{document}